\documentstyle[11pt]{article}
\newlength{\bredde}
\def\slash#1{\settowidth{\bredde}{$#1$}\ifmmode\,\raisebox{.15ex}{/}
\hspace*{-\bredde} #1\else$\,\raisebox{.15ex}{/}\hspace*{-\bredde} #1$\fi}
\newcommand{\be}{\begin{equation}}
\newcommand{\ee}{\end{equation}}
\newcommand{\bea}{\begin{eqnarray}}
\newcommand{\eea}{\end{eqnarray}}

\def\gtwid{\raise.3ex\hbox{$>$\kern-.75em\lower1ex\hbox{$\sim$}}}
\def\ltwid{\raise.3ex\hbox{$<$\kern-.75em\lower1ex\hbox{$\sim$}}}
\begin{document}
\begin{titlepage}
\title{
{\Large Dimensional reduction of dual topological theories}} 

\vspace{0.5cm}

\author{{\sc Kasper Olsen}\thanks{E-mail: kolsen@nbivax.nbi.dk}\\ 
The Niels Bohr Institute\\ Blegdamsvej 17\\ DK-2100 Copenhagen\\ 
Denmark}
\maketitle
\vfill
\begin{abstract} We describe the reduction from four to two dimensions of the 
$SU(2)$ Donaldson-Witten theory and the  dual twisted Seiberg-Witten theory,
i.e. the {\sl Abelian} topological field theory
corresponding to the  
Seiberg--Witten monopole equations.  
\end{abstract}
\vfill
%\vspace{5.5cm}
\begin{flushleft}
NBI-HE-96-14 \\
hep-th/9603023
\end{flushleft}
\end{titlepage}
\newpage
%\phantom{}
%\vfill
%\eject

The recent discovery of duality in $N=2$ supersymmetric Yang-Mills
theory \cite{SW} has a profound impact on the study of topological
invariants of four-manifolds. The field theoretic tool, topological
quantum field theory (for a nice review, see e.g. ref. \cite{danny1}),
was originally based on a non-Abelian gauge theory and associated
ghost systems. This theory, from now on called Donaldson-Witten theory
\cite{D1,W1}, can be obtained from $N=2$ supersymmetric Yang-Mills
theory by an appropriate twisting. If one instead twists the dual
theory, weakly coupled in the infrared, one obtains the same information
from a much simpler Abelian theory. In detail, one studies solutions
to what has become known as the Seiberg-Witten monopole equations
\cite{W2} (for an introduction, see e.g. ref. \cite{matilde}). 
They involve an Abelian gauge potential $A_{\alpha}$ and
a set of commuting Weyl spinors $M$ and $\overline{M}$. The explicit
form of the associated topological Abelian field theory has been
found \cite{labas1}.

Given these recent developments, a natural question to ask is what
happens if one restricts the attention to four-dimensional
product manifolds of the form ${\cal M_4} = M_n\otimes S^{4-n}$
for $n=2, 3$. Taking the radius of the circle $S$ to zero, one
obtains dimensionally reduced theories which by construction are
topological. Such dimensional reductions of Donaldson-Witten theory
have been known for a long time \cite{birm,chap} (and is
briefly reviewed below). As far as topological properties are
concerned, the analogous dimensional reductions of the four-dimensional
{\em dual theory} should provide new Abelian topological theories
which are duals of the dimensionally reduced
Donaldson-Witten theories.

The Donaldson-Witten theory in four dimensions is described by the following
topological action, including a non-Abelian $SU(2)$ gauge potential
$A_{\alpha}$ (see \cite{W1} for details):

\begin{eqnarray}
S^{(4)}&=&\int_{M}d^{4}x\sqrt{g}{\rm Tr}[\frac{1}{4}F_{\alpha\beta}
F^{\alpha\beta}+ \frac{1}{4}F_{\alpha\beta}
\tilde{F}^{\alpha\beta}+\frac{1}{2}\phi D_{\alpha}D^{\alpha}\lambda
-i\eta D_{\alpha}\psi^{\alpha}+2iD_{\alpha}\psi_{\beta}\chi^{\alpha\beta} 
\nonumber \\ 
&&
-\frac{i}{2}\lambda[\psi_{\alpha},\psi^{\alpha}]
-\frac{i}{2}\phi[\eta ,\eta ]-\frac{1}{8}[\phi ,\lambda]^{2}]
\label{dw}
\end{eqnarray}
which is the BRST variation of 

\be
V^{(4)}={\rm Tr}F_{\alpha\beta}^{+}\chi^{\alpha\beta}-
{\rm Tr}\frac{1}{2}B_{\alpha\beta}
\chi^{\alpha\beta}+\frac{1}{2}{\rm Tr}\psi_{\alpha}D^{\alpha}\lambda
-\frac{1}{4}{\rm Tr}(\eta\left[\phi ,\lambda\right] ).
\label{v4}
\ee
$F_{\alpha\beta}^{+}$ is the self--dual part of $F_{\alpha\beta}$, that is
$F_{\alpha\beta}^{+}=\frac{1}{2}(F_{\alpha\beta}+\tilde{F}_{\alpha\beta})$
with $\tilde{F}_{\alpha\beta}=
\frac{1}{2}\epsilon_{\alpha\beta\gamma\delta}F^{\gamma\delta}$. 
The fields transform as
\be
\begin{array}{ll}
\delta A_{\alpha}=i\psi_{\alpha} &  \mbox{} \\
\delta\psi_{\alpha}=- D_{\alpha}\phi &  \mbox{} \\
\delta\phi=0 & \mbox{} \\
\delta\chi_{\alpha\beta}= B_{\alpha\beta} & \delta\lambda=2i\eta \\
\delta B_{\alpha\beta}=0 & \delta\eta=\frac{1}{2}[\phi,\lambda]
\end{array}
\label{t4}
\ee
(the term $\phi\left[ \chi ,\chi\right]$ present in Wittens action \cite{W1} can
be included by adding to eq. (\ref{dw}) a $\delta$-exact term \cite{bau1}).
Note that $\delta^{2}=0$ only up to a gauge transformation,
so $\delta^{2}=0$ on the gauge invariant action $S^{(4)}$. This is in fact enough
to ensure that the theory is topological.
The theory is defined on a four manifold $M$.
Upon dimensionally reducing this action,
we take $M$ to be a product manifold $M=Y\times S^{1}$ with signature
(++++) and assume that all fields are $x^{0}$-independent. Further we
define $\chi^{i}\equiv \chi^{0i}$ such that $\chi^{ij}=\epsilon^{ijk}\chi_{k}$.
This leads to the three-dimensional action ($i,j,k =1,2,3$),

\begin{eqnarray}
S^{(3)}&=&\int_{Y}d^{3}x\sqrt{g}{\rm Tr}[\frac{1}{4}F_{ij}
F^{ij} + \frac{1}{2}F_{ij}\tilde{F}^{ij}
+\frac{1}{2}D_{i}\varphi_{0}D^{i}\varphi_{0}
-\frac{1}{2}[\varphi_{0},\phi ][\varphi_{0},\lambda ]
+\frac{1}{2}\phi D_{i}D^{i}\lambda \nonumber \\
&&-i\eta D_{i}\psi^{i}
-i\eta[\varphi_{0},\psi_{0} ]
+2i\epsilon^{ijk}(D_{i}\psi_{j})\chi_{k}
+2i[\varphi_{0},\psi_{i} ]\chi^{i} +2i\psi_{0}D_{i}\chi^{i}\nonumber \\
&&
-\frac{i}{2}\lambda[\psi_{0},\psi_{0}]
-\frac{i}{2}\lambda[\psi_{i},\psi^{i}]
-\frac{i}{2}\phi[\eta ,\eta ]
-\frac{1}{8}[\phi ,\lambda]^{2}].\nonumber\\
\label{s3}
\end{eqnarray}
where we defined $A_{0}\equiv \varphi_{0}$ and 
$\tilde{F}_{ij}=\epsilon_{ijk}F_{0k}=-\epsilon_{ijk}D_{k}\varphi_{0}$.
The reduction to two dimensions is obtained by assuming that the three
manifold $Y$ is a product  manifold of the form
$Y=X\times S^{1}$ and $x^{1}$-independence of all
fields ($\mu,\nu =2,3$):

\begin{eqnarray}
S^{(2)}&=&\int_{X}d^{2}x\sqrt{g}{\rm Tr}[\frac{1}{4}F_{\mu\nu}F^{\mu\nu} 
+\frac{1}{2}D_{\mu}\varphi_{0}D^{\mu}\varphi_{0}
+\frac{1}{2}D_{\mu}\varphi_{1}D^{\mu}\varphi_{1}
+\frac{1}{2}[\varphi_{1},\varphi_{0}]^{2}\nonumber \\
&&-\frac{1}{2}[\varphi_{0},\phi ][\varphi_{0},\lambda ]
-\frac{1}{2}[\varphi_{1},\phi ][\varphi_{1},\lambda ]
+\frac{1}{2}\phi D_{\mu}D^{\mu}\lambda
-i\eta[\varphi_{0},\psi_{0}] -i\eta[\varphi_{1},\psi_{1}]\nonumber \\
&&-i\eta D_{\mu}\psi^{\mu}
+2i\epsilon^{\mu\nu}(D_{\mu}\psi_{\nu})\chi
+2i\epsilon^{\mu\nu}[\varphi_{1},\psi_{\mu}]\chi_{\nu}
-2i\epsilon^{\mu\nu}(D_{\mu}\psi_{1})\chi_{\nu}\nonumber \\
&&+2i[\varphi_{0},\psi_{1} ]\chi
+2i[\varphi_{0},\psi_{\mu} ]\chi^{\mu}
+2i\psi_{0}D_{\mu}\chi^{\mu}
-2i[\varphi_{1},\psi_{0}]\chi \nonumber \\
&&-
\frac{i}{2}\lambda[\psi_{0},\psi_{0}]
-\frac{i}{2}\lambda[\psi_{1},\psi_{1}]\nonumber \\
&&-\frac{i}{2}\lambda[\psi_{\mu},\psi^{\mu}]
-\frac{i}{2}\phi[\eta ,\eta ]
-\frac{1}{8}[\phi ,\lambda]^{2}\nonumber \\
&&+ \frac{1}{2}[\varphi_{0},\varphi_{1}]\epsilon^{\mu\nu}F_{\mu\nu}
+ \epsilon^{\mu\nu}D_{\mu}\varphi_{1}D_{\nu}\varphi_{0}]
\label{s2}
\end{eqnarray}
where we defined $A_{1}\equiv \varphi_{0}$ and $\chi_{1}\equiv \chi$.
This action is also topological. 
In fact it is the variation of (traces in the adjoint representation implied)

\begin{eqnarray}
V^{(2)}&=&\{F_{\mu\nu}\epsilon_{\mu\nu}
+ 2\left[\varphi_{0},\varphi_{1}\right]\}\chi
-2B\chi+2\{\epsilon_{\mu\nu}D_{\nu}\varphi_{1}- D_{\mu}\varphi_{0}\}\chi_{\mu}
-2B_{\mu}\chi_{\mu} \nonumber\\
&&+\frac{1}{2}\psi_{\mu}D^{\mu}\lambda
+\frac{1}{2}\psi_{0}[\varphi_{0},\lambda ]
+\frac{1}{2}\psi_{1}[\varphi_{1},\lambda ]
-\frac{1}{4}(\eta [\phi ,\lambda ])
\label{v2}
\end{eqnarray}
which is the dimensionally reduced version of the functional $V^{(4)}$ in eq.
(\ref{v4}).
In order to get a simple interpretation of the two-dimensional
topological action, we define
the complex scalar field $\Phi=\varphi_{0}+i\varphi_{1}$.
It is then possible to rewrite the
two-dimensional action as

\begin{eqnarray}
S^{(2)}&=&\int_{X}d^{2}x\sqrt{g}{\rm Tr}
[\frac{1}{4}(F_{\mu\nu}- \frac{1}{2}i\epsilon_{\mu\nu}[\Phi ,\Phi^{*}])^{2}
+\frac{1}{2}D_{\mu}\Phi D_{\mu}\Phi^{*}\nonumber \\
&&-\frac{1}{2}[\varphi_{0},\phi ][\varphi_{0},\lambda ]
-\frac{1}{2}[\varphi_{1},\phi ][\varphi_{1},\lambda ]
+\frac{1}{2}\phi D_{\mu}D^{\mu}\lambda
-i\eta[\varphi_{0},\psi_{0}] -i\eta[\varphi_{1},\psi_{1}]\nonumber \\
&&-i\eta D_{\mu}\psi^{\mu}
+2i\epsilon^{\mu\nu}(D_{\mu}\psi_{\nu})\chi
+2i\epsilon^{\mu\nu}[\varphi_{1},\psi_{\mu}]\chi_{\nu}
-2i\epsilon^{\mu\nu}(D_{\mu}\psi_{1})\chi_{\nu}\nonumber \\
&&+2i[\varphi_{0},\psi_{1} ]\chi
+2i[\varphi_{0},\psi_{\mu} ]\chi_{\mu}
-2i\psi_{0}D_{\mu}\chi^{\mu}
-2i[\varphi_{1},\psi_{0}]\chi\nonumber \\
&&-\frac{i}{2}\lambda[\psi_{0},\psi_{0}]
-\frac{i}{2}\lambda[\psi_{1},\psi_{1}]\nonumber \\
&&-\frac{i}{2}\lambda[\psi_{\mu},\psi_{\mu}]
-\frac{i}{2}\phi[\eta ,\eta ]
-\frac{1}{8}[\phi ,\lambda]^{2}].
\label{s2.2}
\end{eqnarray}
It is easy to check, that the resulting action is a 
BRST gauge fixing of the anti--self-duality
equation in four dimensions, $F_{\alpha\beta}=
- \frac{1}{2}\epsilon_{\alpha\beta\gamma\delta}F^{\gamma\delta}$,
reduced to two dimensions. From the topological point of view, we could
as well have gauge fixed on the self--duality equations
(gauge fixing on $F^{-}$ amounts essentially to extracting from eq. (\ref{dw})
the topological term $\frac{1}{2}\int F\tilde{F}$ and this changes
neither the energy-momentum tensor nor the equations of motion).
These equations have been studied by Hitchin
in ref. \cite{hitchin} which focuses on the topological 
structure of the moduli space
of solutions. Also Chapline and Grossman \cite{chap} 
considered these equations, thereby connecting conformal field theory to 
Donaldson theory.\\

We now turn to the main issue of this letter:
the analogous dimensional reduction of the dual
theory, which is an Abelian gauge theory. 
The fact, that it is Abelian implies enormous simplifications
when it comes to actual calculations 
compared with
its non-Abelian counterpart. 
The theory is defined by a topological action corresponding to
the Seiberg-Witten equations in four dimensions \cite{W2}. These 
equations are
\begin{eqnarray}
F^{+}_{\alpha\beta}&=&-\frac{i}{2}\overline{M}\Gamma_{\alpha\beta}M 
\label{mp}\\
D_{A}M&=& 0 \nonumber
\end{eqnarray}
Here $A$ is an Abelian $U(1)$ gauge potential, 
$D_{A}$ the twisted Dirac 
operator 
and $\Gamma_{\alpha\beta}
=\frac{1}{2}\left[ \gamma_{\alpha}, \gamma_{\beta}\right]$ 
with $\left\{ \gamma_{\alpha} ,\gamma_{\beta}\right\} =2g_{\alpha\beta}$
\footnote{The convention is 
$D_{A}=\gamma^{\alpha}(\partial_{\alpha}+\omega_{\alpha}+iA_{\alpha})$ so 
that $A_{\alpha}$ is real. $\omega_{\alpha}=\frac{1}{8}\omega_{\alpha ij}
[\gamma^{i},\gamma^{j}]$ is the spin connection one-form.}.
$M$ is a (commuting) positive chirality spinor, satisfying $\gamma_{5}M=M$.
As discussed in ref. \cite{labas1,carey} 
one can write a completely 
analogous topological field theory based on these monopole equations
(for a nice review, see e.g. ref. \cite{labas2}).
Let us use the notation of \cite{carey} where the topological action is

\be
S_{m}^{(4)}=\delta V_{m}^{(4)}
\ee
with

\begin{eqnarray}
V_{m}^{(4)}&=&\int_{M}d^{4}x\sqrt{g}\left\{ \left[\partial_{\alpha}\psi^{\alpha}
+\frac{i}{2}(\overline{N}M-\overline{M}N)\right]\lambda
-\chi^{\alpha\beta}(H_{\alpha\beta}
-F^{+}_{\alpha\beta}-\frac{i}{2}\overline{M}\Gamma_{\alpha\beta}M)\right.\nonumber \\
&&\left. -\overline{\mu}(\nu -iD_{A}M)-\overline{(\nu -iD_{A}M)}\mu\right\}.
\end{eqnarray}
The BRST transformation rules are,
\be
\begin{array}{ll}
\delta A_{\alpha}=\psi_{\alpha} &  \mbox{} \\
\delta\psi_{\alpha}=- \partial_{\alpha}\phi &  \delta M=N \\
\delta\phi=0 & \delta N=i\phi M \\
\delta\chi_{\alpha\beta}= B_{\alpha\beta} & \delta\mu=\nu \\
\delta B_{\alpha\beta}=0 & \delta\nu=i\phi\mu \\
\delta\lambda =\eta & \delta\eta=0
\end{array}
\ee
As before $\delta^{2}=0$ up to a gauge transformation.
For instance, $\delta^{2}A_{\alpha}=-\partial_{\alpha}\phi$ which is the
variation of $A_{\alpha}$ under an infinitesimal gauge transformation
generated by $\phi$.
Ghost number assignments are essentially obvious, and can in any case be
found in refs. \cite{labas1,carey}.
Using these transformation rules one finds
the following expression for the topological action
in four dimensions \cite{carey}:

\begin{eqnarray}
S_{m}^{(4)}&=&\int_{M}d^{4}x\sqrt{g}\left\{
\left[ -\Delta\phi+\overline{M}M\phi-i\overline{N}N\right]\lambda
-\left[ \partial_{\alpha}\psi^{\alpha}
+\frac{i}{2}(\overline{N}M-\overline{M}N)\right]\eta
+2i\phi\overline{\mu}\mu\right. \nonumber \\
&&-\chi^{\alpha\beta}\left[ (\partial_{\alpha}\psi_{\beta}
-\partial_{\beta}\psi_{\alpha})^{+}
+\frac{i}{2}(\overline{M}\Gamma_{\alpha\beta}N
+\overline{N}\Gamma_{\alpha\beta}M)\right]\nonumber \\
&&+\frac{1}{4}(F_{\alpha\beta}^{+}
+\frac{i}{2}\overline{M}\Gamma_{\alpha\beta}M)^{2}
+\frac{1}{2}\overline{D_{A}M}D_{A}M\nonumber \\
&&\left. +\overline{(iD_{A}N-\gamma^{\alpha}\psi_{\alpha}M)}\mu
-\overline{\mu}(iD_{A}N-\gamma^{\alpha}\psi_{\alpha}M)\right\}
\end{eqnarray}
where the Lagrange multipliers $B$ and $\nu$ 
are eliminated by their equations of motion and the bar indicates
hermitian conjugation.
By taking $M=Y\times S^{1}$ the dimensionally reduced action becomes:

\begin{eqnarray}
S_{m}^{(3)}&=&\int_{Y}d^{3}x\sqrt{g}\left\{
\left[ -\Delta\phi+\overline{M}M\phi-i\overline{N}N\right]\lambda
-\left[ \partial_{k}\psi^{k}
+\frac{i}{2}(\overline{N}M-\overline{M}N)\right]\eta
+2i\phi\overline{\mu}\mu\right. \nonumber \\
&&-2\chi^{k}\left[ -\partial_{k}\psi_{0}+\epsilon_{kij}(\partial_{i}\psi_{j})
-\overline{M}\sigma_{k}N
-\overline{N}\sigma_{k}M\right]\nonumber \\
&&+\frac{1}{8}(F_{ij}-\epsilon_{ijk}\partial_{k}\varphi_{0}
-\epsilon_{ijk}\overline{M}\sigma_{k}M)^{2}
+\frac{1}{2}\overline{(D_{A}+\varphi_{0})M}(D_{A}+\varphi_{0})M\nonumber \\
&&+\left. \overline{( i(D_{A}+\varphi_{0})N-(\sigma^{k}\psi_{k}-i\psi_{0})M)}\mu
-\overline{\mu}( i(D_{A}+\varphi_{0})N-(\sigma^{k}\psi_{k}-i\psi_{0})M)\right\}
\label{act3}
\end{eqnarray}
where $\chi^{i}\equiv \chi^{0i}$ and $\sigma_{k}$ are the Pauli matrices.
This dimensionally reduced action was obtained in ref. \cite{carey}. 
The partition function of this three dimensional theory gives a Seiberg-Witten
version of the Casson invariant \cite{carey}.
The monopole equations (\ref{mp}) reduce, as follows from eq. (\ref{act3}), 
to the
following set of equations:

\begin{eqnarray}
F_{ij}-\epsilon_{ijk}\overline{M}\sigma_{k}M &=& 0 \nonumber\\
D_{A}M &=& 0 \label{mp3}\\
\varphi_{0} &=& 0.\nonumber
\end{eqnarray}
The last condition is only necessary if we have a non--trivial solution.
If not, it can be replaced by the condition $d\varphi_{0}=0$.
Finally, we make a reduction to two dimensions 
(with $Y=X\times S^{1}$ ) which results in the following action:

\begin{eqnarray}
S_{m}^{(2)}&=&\int_{X}d^{2}x\sqrt{g}\left\{
\left[ -\Delta\phi+\overline{M}M\phi-i\overline{N}N\right]\lambda
-\left[ \partial_{\mu}\psi^{\mu}
+\frac{i}{2}(\overline{N}M-\overline{M}N)\right]\eta
+2i\phi\overline{\mu}\mu\right. \nonumber \\
&&-2\chi^{\mu}\left[ -\partial_{\mu}\psi_{0}
+\epsilon_{\mu\nu}\partial_{\nu}\psi_{1}
-\overline{M}\sigma_{\mu}N
-\overline{N}\sigma_{\mu}M\right]
-2\chi\left[ \epsilon_{\mu\nu}(\partial_{\mu}\psi_{\nu})
-\overline{M}\sigma_{1}N-\overline{N}\sigma_{1}M\right]\nonumber \\
&&+\frac{1}{8}(F_{\mu\nu}
-\epsilon_{\mu\nu}\overline{M}\sigma_{1}M)^{2}
+\frac{1}{4}\partial_{\mu}\varphi_{0}\partial_{\mu}\varphi_{0}
+\frac{1}{4}\partial_{\mu}\varphi_{1}\partial_{\mu}\varphi_{1}
%+\frac{1}{2}\epsilon_{\mu\nu}\partial_{\mu}\varphi_{1}
%\partial_{\nu}\varphi_{0}
\nonumber\\
&&+\frac{1}{2}\partial_{\mu}\varphi_{1}\epsilon_{\mu\nu}
\overline{M}\sigma_{\nu}M
+\frac{1}{2}\partial_{\mu}\varphi_{0}\overline{M}\sigma_{\mu}M
+\frac{1}{4}(\overline{M}\sigma_{\mu}M)^{2}
\nonumber \\
&&+\frac{1}{2}\overline{(D_{A}+\varphi_{0}+i\varphi_{1}\sigma_{1})M}
(D_{A}+\varphi_{0}+i\varphi_{1}\sigma_{1})M\nonumber \\
&& +\overline{( i(D_{A}+\varphi_{0}+i\varphi_{1}\sigma_{1})N-(\sigma^{\mu}\psi_{\mu}
-i\psi_{0}+\psi_{1}\sigma_{1})M)}\mu\nonumber \\
&&\left. -\overline{\mu}( i(D_{A}+\varphi_{0}+i\varphi_{1}\sigma_{1})N
-(\sigma^{\mu}\psi_{\mu}-i\psi_{0}+\psi_{1}\sigma_{1})M)\right\}
\end{eqnarray}
One can check that the resulting two-dimensional action also is 
topological since 
$S_{m}^{(2)}=\delta V_{m}^{(2)}$. Here $V_{m}^{(2)}$ is the
dimensional reduction of $V_{m}^{(4)}$, i.e.

\begin{eqnarray}
V_{m}^{(2)}&=&\int_{X}d^{2}x\sqrt{g}\left\{ \left[\partial_{\mu}\psi^{\mu}
+\frac{i}{2}(\overline{N}M-\overline{M}N)\right]\lambda
-2\chi(2H
-\frac{1}{2}\epsilon_{\mu\nu}F_{\mu\nu}+\overline{M}\sigma_{1}M)
\right.\nonumber \\
&&-2\chi_{\mu}(2H_{\mu}+\epsilon_{\mu\nu}\partial_{\nu}\varphi_{1}
-\partial_{\mu}\varphi_{0}-\overline{M}\sigma_{\mu}M)\nonumber \\
&&\left. -\overline{\mu}(\nu 
-i(D_{A}+\varphi_{0}+i\varphi_{1}\sigma_{1})M)-
\overline{(\nu -i(D_{A}+\varphi_{0}+i\varphi_{1}\sigma_{1})M)}\mu\right\}.
\end{eqnarray}
Here $\chi\equiv\chi^{1}$, $H^{\mu}\equiv H^{0\mu}$ and $H\equiv H^{1}$. 
The monopole equations are inferred from the classical part of the
action. It is

\begin{eqnarray}
S_{0}&=&\frac{1}{8}(F_{\mu\nu}-\epsilon_{\mu\nu}\overline{M}
\sigma_{1}M)^{2}+\frac{1}{4}(\partial_{\mu}\varphi_{1})^{2}
+\frac{1}{2}(\partial_{\mu}\varphi_{0})^{2}
+\frac{1}{4}(\overline{M}\sigma_{\mu}M)^{2}\nonumber \\
&&+\frac{1}{2}|D_{A}M|^{2}+\frac{1}{2}|\varphi_{0}M|^{2}
+\frac{1}{2}|\varphi_{1}\sigma_{1}M|^{2}
\end{eqnarray}
In two dimensions the monopole equations therefore reduce to 
 
\begin{eqnarray}
F_{\mu\nu}-\epsilon_{\mu\nu}\overline{M}\sigma_{1}M &=& 0 \nonumber\\
D_{A}M &=& 0 \nonumber\\
\overline{M}\sigma_{\mu}M &=& 0\label{unus}\\
\varphi_{0} &=& 0 \nonumber\\
\varphi_{1} &=& 0.\nonumber
\end{eqnarray}
If $(A,M)$ is a trivial solution, then the last two conditions
can be replaced by $d\varphi_{0}=d\varphi_{1}=0$.
Due to the consistency of the monopole equations in four, three and two
dimensions, the first of the above mentioned 
relations could as well have been deduced
directly from the monopole equations (\ref{mp}) in four dimensions.
For example, consider the $F^{+}_{01}$ component of eq. (\ref{mp}). By 
dimensional reduction we get $\frac{1}{4}\epsilon_{\mu\nu}F^{\mu\nu}$
on the l.h.s and $\frac{1}{2}\overline{M}\sigma_{1}M$ on the r.h.s.,
which is the wanted relation. 
Similarly, the unusual-looking condition (\ref{unus}) is nothing but
one of the dimensionally reduced remnants of the 3-d monopole eqs. 
(\ref{mp3}) and can be derived from the $F_{1\mu}$ part of the first of
these equations together with $\varphi_{1}=0$.
It is possible to take $M$ to be a positive chirality spinor ($\sigma_{1}M=M$),
but this is not in it self dictated by the dimensional reduction. 

Some explicit solutions to 
the monopole equations on ${\bf R}^{2}$ have been constructed 
in ref. \cite{nergiz}
and the solutions turn out to be vortex configurations.
They are singular, as are the analogous 3-d solutions given 
by Freund in ref. \cite{freund}. As noted by Witten in \cite{W2},
the monopole equations admit no $L^{2}$ solutions on flat 
${\bf R}^{n}, n\leq4$.

The vanishing theorems of Witten \cite{W2} also have an analogous formulation
in two dimensions, as noted by Thompson in ref. \cite{thompson1}
where a brief discussion of the dimensional reduction of the 
monopole equations also can be found. 
In fact, it follows from eq. (\ref{unus}), that if
$(A,M)$ is a solution of the two-dimensional monopole equations 
then the pair must obey the following identity

\be
\int_{X}d^{2}x\sqrt{g}(\frac{1}{4}|F|^{2}+\overline{D^{\mu}M}D_{\mu}M+
\frac{1}{2}|\overline{M}\sigma_{1}M|^{2}+\frac{1}{4}R|M|^{2})=0
\ee
where $R$ is the scalar curvature.
If there is a metric so that $R$ is positive on $X$ then this would imply
that $F_{\mu\nu}=0$ and $M=0$ are the only solutions. On a sphere,
for example, we are actually looking at flat Abelian connections.\\ 
%(In the general case one finds the inequality 

%\be
%\int_{X}d^{2}x\sqrt{g}\frac{1}{4}|F|^{2} 
%\leq \frac{1}{32}\int_{X}d^{2}x\sqrt{g}R^{2}.)
%\ee

In summary, we have found an expression for the topological action
corresponding to the reduction of the monopole equations to
two dimensions. As far as topological properties are concerned, 
this Abelian theory should be equivalent (dual) to the corresponding
reduction of the Donaldson-Witten theory. Therefore one could 
speculate whether it is possible to extract from the Abelian theory
information about the moduli space of solutions to the
self--duality equations on a Riemann surface studied by Hitchin
\cite{hitchin}. 
In connection with this, we may also refer to the work of Brooks and
Lue \cite{brooks}, where it is shown that solutions to a 
twisted version of the
four dimensional monopole equations are realized as a subset of 
solutions to the Donaldson-Witten theory. One may wonder whether a
corresponding relation exists between the theory of Hitchin and
the analogous twisting of the two dimensional monopole equations
suggested here.
It is intriguing to note that 
the two--dimensional topological theory corresponding
to the monopole equations is very similar to the  
two--dimensional
topological Abelian--Higgs model of Schaposnik and Thompson \cite{schapos}.
It is of interest to determine if they are in fact directly related.
We hope to return to these questions in a future publication.\\

{\sc Acknowledgement:}~
The author would like to thank J.M.F. Labastida for helpful discussions and
remarks.

%\newpage

\end{document}